\documentclass[aps,prb,showpacs,amsmath,twocolumn]{revtex4}%,preprint
\usepackage{bm}
\usepackage{graphicx}

\begin{document}

\title{Analytical model of the response of a superconducting film to line currents}

\author{Yasunori Mawatari}
% \email[]{y.mawatari@aist.go.jp}
\affiliation{%
	National Institute of Advanced Industrial Science and Technology (AIST), 
	Tsukuba, Ibaraki 305--8568, Japan}

\author{John R. Clem}
\affiliation{%
	Ames Laboratory--DOE and Department of Physics and Astronomy, 
	Iowa State University, Ames Iowa 50011, USA}

\received{July 3, 2006}
\revised{September 15, 2006}
% \received{}\revised{}\accepted{}\published{}

\begin{abstract}
We theoretically investigate the response of a superconducting film to 
line currents flowing in linear wires placed above the film, 
and we present analytic expressions for the magnetic-field and current 
distributions based on the critical state model. 
The behavior of the superconducting film is characterized by the 
sheet-current density $K_z$, whose magnitude cannot exceed the critical 
value $j_cd$, where $j_c$ is the critical current density and $d$ is 
the thickness of the film. 
When the transport current $I_0$ flowing in the wire is small enough, 
$|K_z|$ is smaller than $j_cd$ and the magnetic field is shielded 
below the film. 
When $I_0$ exceeds a threshold value $I_{c0}\propto j_cd$, 
on the other hand, $|K_z|$ reaches $j_cd$ and the magnetic field 
penetrates below the film. 
We also calculate the ac response of the film when an ac transport 
current flows in the linear wires. 
\end{abstract}

% 74.25.Nf	Response to electromagnetic fields
% 74.25.Sv	Critical currents 
% 74.78.-w	Superconducting films and low-dimensional structures 

\maketitle

\section{Introduction%===========================
\label{Sec_Intro}}
The response of superconducting films to homogeneous applied magnetic 
fields is well understood, and analytic expressions for the distributions 
of the magnetic field and current density have been 
derived~\cite{Swan68,Halse70,Mikheenko93,Zhu93,Brandt93,Zeldov94a} 
based on Bean's critical state model.~\cite{Bean62} 
When small current-carrying coils are placed near the surface to probe 
the local properties of superconducting films, the magnetic fields 
generated by the coils are inhomogeneous. 
Analytic expressions describing the response of superconducting films 
to small coils have been derived for the linear response 
regime,~\cite{Fiory88,Clem92,Klupsch95,Gilchrist96,Turneaure98,Coffey01} 
but in order to measure the local distribution of the critical current 
density $j_c$ in superconducting films, it is necessary to investigate 
the nonlinear 
response.~\cite{Bernhardt89,Claassen91,Poulin93,Hochmuth94,Mawatari02} 
Numerical computations of the nonlinear response of superconducting films 
to the inhomogeneous magnetic fields arising from small coils have been 
carried out in 
Refs.~\onlinecite{Gilchrist96,Koo96,Wada03,Yamada05,Aurino05}, 
but here we present analytic results for the nonlinear response 
to line currents above superconducting films. 

The procedure proposed by Claassen {\it et al.}~\cite{Claassen91} 
for inductive measurements of the local $j_c$ distribution in films 
of thickness much greater than the London penetration depth $\lambda$ 
can be described briefly as follows. 
A small coil carrying a sinusoidal drive current $I_0\cos\omega t$ is 
placed just above a superconducting film, and the induced voltage 
$V(t)= \sum_{n} V_n\cos(n\omega t+\vartheta_n)$ in the coil is detected. 
The amplitude of the third-harmonic voltage $V_3$ is measured as a 
function of the drive current amplitude $I_0$, and the threshold current 
$I_{c0}$ is defined such that $V_3=0$ for $0<I_0<I_{c0}$ and $V_3>0$ 
for $I_0>I_{c0}$. 
Because $I_{c0}\propto j_cd$, where $d$ the film thickness, $j_c$ 
can be evaluated from $I_{c0}$.~\cite{Claassen91,Poulin93,Mawatari02} 
Since an electric-field criterion must be applied for a precise 
determination of $j_c$, it is important to evaluate the electric field 
$E_f$ generated in the superconducting film.~\cite{Yamasaki03} 

In the present paper we consider linear wires as simple models of 
coil wires, and we analytically investigate the response of 
a superconducting film to linear wires carrying transport currents. 
In Sec.~\ref{Sec_single-wire} we investigate the dc (ac) response of a 
superconducting film to a linear wire carrying a dc (ac) transport current: 
we determine the threshold current $I_{c0}$, and we present the voltage 
$V(t)$ and the harmonic voltages induced in the linear wire, 
as well as the electric field $E_f$ induced in the superconducting film. 
In Sec.~\ref{Sec_two-wires} we consider a superconducting film and two 
linear wires carrying transport currents of opposite directions. 
We briefly summarize our results in Sec.~\ref{Sec_conclusion}.

\section{%=====================================================
A single linear wire and a superconducting film
\label{Sec_single-wire}}

\begin{figure}[b]%*************
 \includegraphics{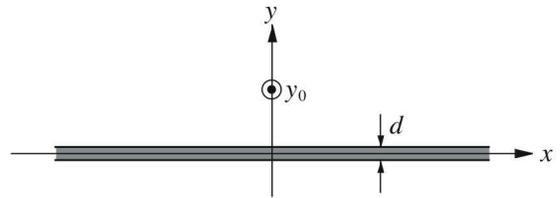}
\caption{%
Configuration of a superconducting film at $y=0$ and a linear wire 
at $(x,y)=(0,y_0)$. 
The film is infinitely extended in the $xz$ plane, and the infinite wire 
is parallel to the $z$ axis. 
}
\label{Fig_SC-wire}
\end{figure}

In this section we consider a superconducting film and a linear wire 
carrying a transport current, as shown in Fig.~\ref{Fig_SC-wire}. 
An infinitely long wire, parallel to the $z$ axis, is situated at 
$(x,y)=(0,y_0)$ where $y_0>0$. 
The radius of the wire $r_w$ is assumed to be much smaller than $y_0$. 
A superconducting film, infinitely extended in the $xz$ plane, is 
situated at $-d/2<y<+d/2$, where the film thickness $d$ is much smaller 
than $y_0$ but is larger than the London penetration depth $\lambda$. 
Flux pinning in the film is characterized by the critical current density 
$j_c$, which is assumed to be constant (independent of magnetic field) 
as in Bean's critical state model~\cite{Bean62} 
and to be spatially homogeneous in the film. 
We consider the limit $d\to 0$, as this simplification allows us to 
obtain simple analytic expressions for the magnetic-field and current 
distributions. 
In the thin-film limit of $d\to 0$, the sheet current 
$K_z(x)=\int_{-d/2}^{+d/2}j_z(x,y)dy$ plays crucial roles, 
and the upper limit of $|K_z|$ is the critical sheet-current density $j_cd$. 
The lower critical field $H_{c1}$ is assumed to be much smaller than 
$j_cd$ (i.e., $H_{c1}/j_cd\to 0$), such that the flux penetration into 
superconducting films is not affected by $H_{c1}$, but is determined by 
$j_cd$.~\cite{Clem06} 

We introduce the complex field 
${\cal H}(\zeta)=H_y(x,y)+iH_x(x,y)$,~\cite{Clem73,Zeldov94b,Mawatari01} 
which is an analytic function of $\zeta=x+iy$ for $y\neq 0$ and 
$(x,y)\neq (0,y_0)$. 
The Biot-Savart law for the complex field is given by 
\begin{equation}
	{\cal H}(\zeta)= {\cal H}_0(\zeta) +\frac{1}{2\pi} 
		\int_{-\infty}^{+\infty}du \frac{K_z(u)}{\zeta -u} , 
\label{Biot-Savart}
\end{equation}
where ${\cal H}_0(\zeta)$ is the complex field arising from the line 
current alone. 
The ${\cal H}_0(\zeta)$ is given by 
\begin{equation}
	{\cal H}_0(\zeta) =\frac{I_z}{2\pi} \frac{1}{\zeta-iy_0} , 
\label{h0}
\end{equation}
where $I_z$ is the transport current flowing in the linear wire. 
At the upper ($\zeta=x+i\epsilon$) and lower ($\zeta=x-i\epsilon$) 
surfaces of the superconducting film, where $\epsilon =d/2$ is a positive 
infinitesimal, the perpendicular and parallel magnetic fields 
$H_y(x,0)=\mbox{Re}\,{\cal H}(x\pm i\epsilon)$ and 
$H_x(x,\pm \epsilon)=\mbox{Im}\,{\cal H}(x\pm i\epsilon)$ are obtained 
from Eq.~\eqref{Biot-Savart} as 
\begin{eqnarray}
	H_y(x,0) &=& \mbox{Re}\,{\cal H}_0(x) +\frac{1}{2\pi} 
		\mbox{P} \int_{-\infty}^{+\infty}du \frac{K_z(u)}{x-u} , 
\label{Hy_x+-i0}\\
	H_x(x,\pm \epsilon) &=& \mbox{Im}\,{\cal H}_0(x)\mp K_z(x)/2 , 
\label{Hx_x+-i0}
\end{eqnarray}
where P denotes the Cauchy principal value integral. 
The complex potential is defined by 
${\cal G}(\zeta)=\int{\cal H}(\zeta)d\zeta$, and the contour lines 
of the real part of ${\cal G}(\zeta)$ correspond to magnetic-field lines. 

The magnetic flux per unit length $\Phi_w$ around the linear wire is 
\begin{eqnarray}
	\Phi_w &=& -\mu_0\int_{y_0+r_0}^{\infty} dy\,H_x(0,y) 
\nonumber\\
	&=& \mu_0\mbox{Re}\,\left[ {}-{\cal G}\bigl(i(y_0+r_0)\bigr) 
		+\lim_{v\to\infty} {\cal G}(iv) \right] . 
\label{Phi-wire_def}
\end{eqnarray}
We have introduced a cutoff length $r_0\ll y_0$, where $r_0$ is of the 
order of the radius of the wire, to remove the logarithmic divergence 
as $\zeta\to iy_0$. 
The magnetic flux per unit length $\Phi_f(x')$ up through the film 
($y=0$) in the region $x'<x<+\infty$ is 
\begin{eqnarray}
	\Phi_f(x') &=& \mu_0\int_{x'}^{\infty} du\,H_y(u,0) 
\nonumber\\
	&=& \mu_0\mbox{Re}\,\left[ {}-{\cal G}(x') 
		+\lim_{u\to\infty} {\cal G}(u) \right] . 
\label{Phi-film_def}
\end{eqnarray}

\subsection{DC response%**************************
\label{Sec_DC}}
In this subsection we consider the magnetic-field distribution 
when the linear wire carries a dc current $I_z=I_0>0$ that is held 
constant after monotonically increasing from $I_z=0$. 

\subsubsection{Linear response for $0<I_0 \le I_{c0}$%=====================
\label{Sec_DC-linear}}
\begin{figure*}[bt]%*************
 \includegraphics{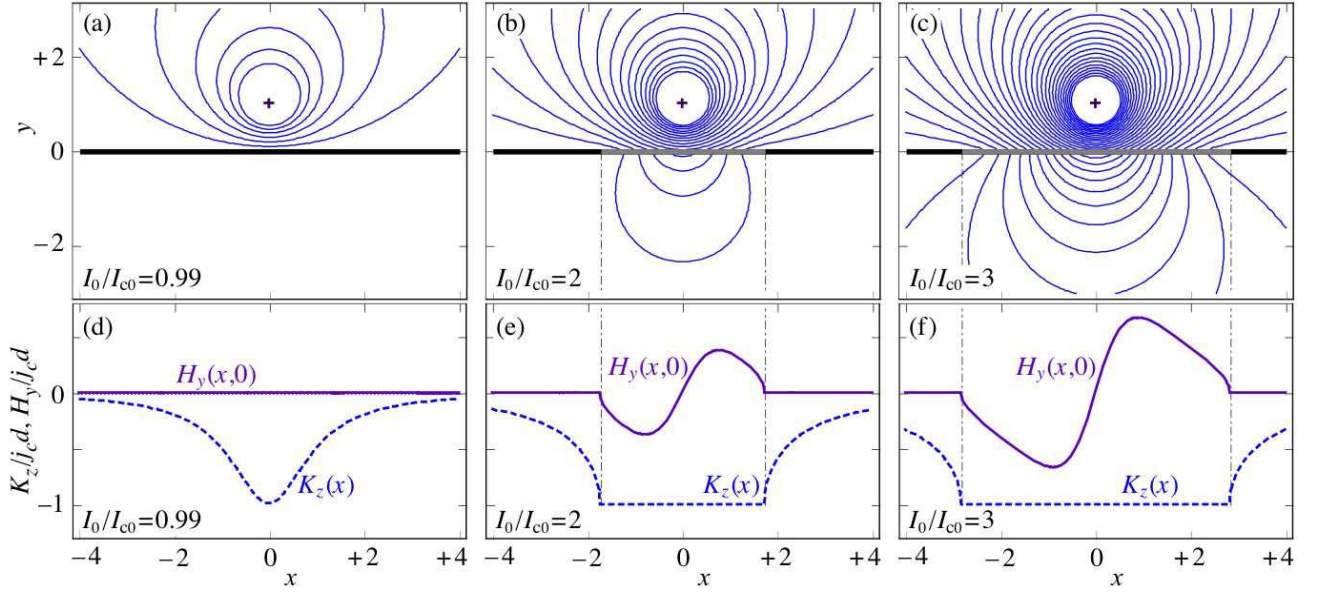}
\caption{(color online) 
Magnetic-field and sheet-current distributions 
(a,d) for $I_0/I_{c0}=0.99$, (b,e) for $I_0/I_{c0}=2$ $(a=1.73)$, 
and (c,f) for $I_0/I_{c0}=3$ $(a=2.83)$. 
Top figures (a,b,c) show the magnetic-field lines around a linear wire 
and a superconducting film. 
The cross symbols at $(x,y)=(0,y_0)=(0,1)$ denote the position of 
the linear wire, the black horizontal line the shielded region $|x|>a$ 
of the superconducting film, and the gray horizontal lines the penetrated 
region $|x|<a$ of the film. 
The vertical dot-dashed lines indicate the positions of the flux front 
at $x=\pm a$. 
Bottom figures (d,e,f) show the distributions of the perpendicular 
magnetic field $H_y(x,0)$ divided by $j_cd$ (solid lines) 
and the sheet-current density $K_z(x)$ divided by $j_cd$ (dashed lines) 
in the superconducting film. 
}
\label{Fig_field-lines}
\end{figure*}

For $0<I_0 \le I_{c0}$, the magnetic field is completely shielded 
below the film, $y=\mbox{Im}\,\zeta <0$. 
The field distribution can be obtained by the mirror-image technique, 
and the resulting complex field is 
\begin{equation}
	{\cal H}(\zeta)= 
	\begin{cases} \displaystyle 
		\frac{I_0}{\pi}\frac{iy_0}{\zeta^2+y_0^2} 
		& \mbox{for } \mbox{Im}\,\zeta >0 , \\
		0 & \mbox{for } \mbox{Im}\,\zeta <0 . 
	\end{cases}
\label{cH_linear}
\end{equation}
The complex potential ${\cal G}(\zeta)=\int{\cal H}(\zeta)d\zeta$ 
for $\mbox{Im}\,\zeta >0$ is given by 
\begin{eqnarray}
	{\cal G}(\zeta) = 
		\frac{I_0}{\pi}i\arctan\left(\frac{\zeta}{y_0}\right) . 
\label{cG_linear}
\end{eqnarray}
The perpendicular magnetic field and sheet-current density are thus 
given by $H_y(x,0)=0$ and 
\begin{equation}
	K_z(x) = -\frac{I_0}{\pi} \frac{y_0}{x^2+y_0^2} , 
\label{Kz_linear}
\end{equation}
respectively. 
The net current induced in the superconducting film is 
$\int_{-\infty}^{+\infty}K_z(x)dx =-I_0$, as expected. 
Note that the sheet-current density cannot exceed $j_cd$; 
that is, $|K_z(x)|\leq j_cd$. 
Because the maximum of $|K_z|$ given by Eq.~\eqref{Kz_linear} 
is $I_0/\pi y_0$, Eq.~\eqref{Kz_linear} is valid for $I_0\leq I_{c0}$, 
where the threshold current is given by 
\begin{equation}
	I_{c0}= \pi j_c d y_0 . 
\label{Ic0}
\end{equation}
Figure~\ref{Fig_field-lines}(a) shows the magnetic-field lines 
[i.e., the contour lines of $\mbox{Re}\,{\cal G}(x+iy)$] calculated 
from Eq.~\eqref{cG_linear}, and the dashed line in 
Fig.~\ref{Fig_field-lines}(d) shows $K_z(x)$ given by Eq.\eqref{Kz_linear}. 

%*************

The magnetic flux per unit length around the linear wire, calculated by 
substituting Eq.~\eqref{cH_linear} into Eq.~\eqref{Phi-wire_def}, is 
\begin{equation}
	\Phi_w= \frac{\mu_0I_0}{\pi} \int_{y_0+r_0}^{\infty} dy\, 
		\frac{y_0}{y^2-y_0^2} 
	= L_0I_0 , 
\label{Phi-wire_linear}
\end{equation}
where the inductance per unit length $L_0=(\mu_0/2\pi)\ln(2y_0/r_0)$ 
corresponds to the difference between the self inductance of the linear 
wire and the mutual inductance of the wire and its image. 
Because the perpendicular magnetic field in the film is zero, 
the magnetic flux up through the film defined by Eq.~\eqref{Phi-film_def} 
is also zero:~\cite{foot0} 
\begin{equation}
	\Phi_f(x)= 0 . 
\label{Phi-film_linear}
\end{equation}

Equations~\eqref{cH_linear}, \eqref{cG_linear}, \eqref{Kz_linear}, 
\eqref{Phi-wire_linear}, and \eqref{Phi-film_linear} are valid 
for $0<I_0 \le I_{c0}$. 

\subsubsection{Nonlinear response for $I_0>I_{c0}$%==========================
\label{Sec_DC-nonlinear}}
For $I_0>I_{c0}$, on the other hand, the maximum $|K_z|$ reaches 
$j_cd$ and the magnetic field penetrates below the superconducting film. 
The field distributions for $I_0>I_{c0}$, therefore, must satisfy 
\begin{eqnarray}
	K_z(x)= -j_cd \ \mbox{ and } H_y(x,0)\neq 0 
	&&\mbox{for } |x|<a, 
\label{KzHy_nonlinear_x<a}\\
	|K_z(x)|< j_cd \ \mbox{ and } H_y(x,0) =0 
	&&\mbox{for } |x|>a, 
\label{KzHy_nonlinear_x>a}
\end{eqnarray}
where the flux fronts [i.e., the boundaries between the region of 
$H_y(x,0)\neq 0$ and that of $H_y(x,0)=0$] are at $x=\pm a$. 
The complex field and the parameter $a$ are determined such that they 
are consistent with Eqs.~\eqref{KzHy_nonlinear_x<a} and 
\eqref{KzHy_nonlinear_x>a}, as derived in Appendix \ref{App_deriv-cH}. 
The parameter $a$ is a function of $I_0$, 
\begin{equation}
	a= y_0\sqrt{(I_0/I_{c0})^2 -1} , 
\label{a-I0}
\end{equation}
and the complex field is 
\begin{equation}
	{\cal H}(\zeta)= \frac{j_cd}{2} 
		\left(\pm i +\frac{iy_0\sqrt{a^2+y_0^2} 
		+\zeta\sqrt{a^2-\zeta^2}}{\zeta^2+y_0^2} \right) , 
\label{cH_nonlinear}
\end{equation}
where the upper sign holds for Im$\zeta>0$ and the lower sign for Im$\zeta<0$. 
The corresponding complex potential is 
\begin{eqnarray}
	{\cal G}(\zeta) &=& \frac{j_cd}{2} 
		\left\{{}\pm i\zeta +\sqrt{a^2-\zeta^2} 
		+\sqrt{a^2+y_0^2} 
		\phantom{\sqrt{\frac{\zeta^2}{y_0^2}}} \right. 
\nonumber\\
	&& \left.{}\times \left[ i\arctan\left(\frac{\zeta}{y_0}\right) 
		-\mbox{arctanh}\left(\sqrt{%
		\frac{a^2-\zeta^2}{a^2+y_0^2}}\right) \right]\right\} . 
\nonumber\\
\label{cG_nonlinear}
\end{eqnarray}
Figures~\ref{Fig_field-lines}(b) and (c) show the magnetic-field lines, 
i.e., the contour lines of $\mbox{Re}\,{\cal G}(x+iy)$ calculated 
from Eq.~\eqref{cG_nonlinear}. 

The perpendicular magnetic field 
$H_y(x,0)=\mbox{Re}\,{\cal H}(x\pm i\epsilon)$ in the superconducting film, 
obtained from Eq.~\eqref{cH_nonlinear}, is 
\begin{equation}
	H_y(x,0)= 
	\begin{cases}
		\displaystyle \frac{j_cd}{2}\frac{x\sqrt{a^2-x^2}}{x^2+y_0^2} 
		& \mbox{for } |x|<a , \\[1ex]
		0 & \mbox{for } |x|>a , 
	\end{cases}
\label{Hy_nonlinear}
\end{equation}
and the sheet-current density 
$K_z(x)=\mbox{Im}\,[{\cal H}(x-i\epsilon)-{\cal H}(x+i\epsilon)]$ is 
\begin{eqnarray}
	K_z(x) &=& 
	\begin{cases}
		-j_cd & \mbox{for } |x|<a , \\[1ex]
		\displaystyle -j_cd \left(1 
		-\frac{|x|\sqrt{x^2-a^2}}{x^2+y_0^2} \right) 
		& \mbox{for } |x|>a . 
	\end{cases}
\nonumber\\
\label{Kz_nonlinear}
\end{eqnarray}
The net current induced in the superconducting film is again 
$\int_{-\infty}^{+\infty}K_z(x)dx =-I_0$, as expected. 
Figures~\ref{Fig_field-lines}(e) and (f) show $H_y(x,0)$ and $K_z(x)$ 
given by Eqs.~\eqref{Hy_nonlinear} and \eqref{Kz_nonlinear}, respectively. 

\begin{widetext}%**************************** 
The magnetic flux per unit length around the linear wire, calculated by 
substituting Eq.~\eqref{cG_nonlinear} into Eq.~\eqref{Phi-wire_def}, is 
\begin{equation}
	\Phi_w \simeq -\frac{\mu_0j_cd}{2} 
		\left[ y_0-\sqrt{a^2+y_0^2} +\sqrt{a^2+y_0^2} 
		\ln\left(\frac{2\sqrt{a^2+y_0^2}}{r_0}\right)\right] . 
\label{Phi-wire_nonlinear}
\end{equation}
Combining Eqs.~\eqref{Phi-wire_linear}, \eqref{a-I0}, and 
\eqref{Phi-wire_nonlinear} yields $\Phi_w=\widetilde{\Phi}_w(I_0)$, where 
\begin{eqnarray}
	\widetilde{\Phi}_w(I_0) &=& 
	\begin{cases}
		L_0I_0 
		& \mbox{for } 0<I_0 \le I_{c0} , \\[2ex]
		\displaystyle
		L_0I_0 +\frac{\mu_0}{2\pi} \left[ I_{c0}-I_0+I_0 
		\ln\left(\frac{I_0}{I_{c0}}\right)\right] 
		& \mbox{for } I_0>I_{c0} . 
	\end{cases}
\label{Phi_dc-def}
\end{eqnarray}
The magnetic flux up through the film, calculated by substituting 
Eq.~\eqref{cG_nonlinear} into Eq.~\eqref{Phi-film_def}, is 
\begin{equation}
	\Phi_f(x) 
	= \frac{\mu_0j_cd}{2} 
		\left[-\sqrt{a^2-x^2} +\sqrt{a^2+y_0^2}\, 
		\mbox{arctanh} \left( \sqrt{\frac{a^2-x^2}{a^2+y_0^2}}\, 
		\right)\right] 
\label{Phi-film_nonlinear}
\end{equation}
for $|x|<a$, and $\Phi_f(x)=0$ for $|x| \ge a$. 
Combining Eqs.~\eqref{Phi-film_linear}, \eqref{a-I0}, 
and \eqref{Phi-film_nonlinear} yields 
$\Phi_f(x)=\widetilde{\Phi}_f(x,I_0)$, where 
\begin{eqnarray}
	\widetilde{\Phi}_f(x,I_0) &=& 
	\begin{cases}
		0 & \mbox{for } 0<I_0 \le I_{c0} \mbox{ or } |x| \ge a , \\[2ex]
		\displaystyle
		\frac{\mu_0}{2\pi} \left[ 
		\sqrt{I_0^2 -I_{c0}^2\left(1+\frac{x^2}{y_0^2}\right)} 
		+I_0\mbox{arccosh} \left(\frac{I_0}{I_{c0}} 
		\frac{y_0}{\sqrt{x^2+y_0^2}}\right) \right] 
		& \mbox{for } I_0>I_{c0} \mbox{ and } |x|<a. 
	\end{cases}
\label{Phi-film_dc-def}
\end{eqnarray}
\end{widetext}%****************************

\subsection{AC response%**************************
\label{Sec_AC}}
In this subsection we consider the time-dependent field distributions 
when the linear wire carries a sinusoidal ac drive current 
$I_z(t)=I_0\cos\omega t$. 
In inductive measurements of the local $j_c$ in superconducting films, 
harmonic voltages induced in coils are 
detected.~\cite{Claassen91,Poulin93,Mawatari02} 
For precise $j_c$ measurements, it also is important to determine 
the electric field induced in the film.~\cite{Yamasaki03} 

We wish to calculate the magnetic field around the linear wire 
$\Phi_w(t)$, defined by Eq.~\eqref{Phi-wire_def}, and the voltage 
per unit length induced in the wire, 
\begin{equation}
	V(t)= R_wI_0\cos\omega t -d\Phi_w(t)/dt , 
\label{V-wire_def}
\end{equation}
where $R_w$ is the resistance per unit length of the wire. 
We also wish to determine the magnetic flux per unit length up through 
the film $\Phi_f(x,t)$, defined by Eq.~\eqref{Phi-film_def}, 
and the electric field in the film,\cite{foot1} 
\begin{equation}
	E_f(x,t)= -\partial\Phi_f(x,t)/\partial t . 
\label{E-film_def}
\end{equation}

\subsubsection{Linear response for $0<I_0 \le I_{c0}$%========================
\label{Sec_AC-linear}}
For $0<I_0 \le I_{c0}$ the magnetic field is completely shielded 
below the superconducting film, $y=\mbox{Im}\,\zeta <0$, 
as in Sec.~\ref{Sec_DC-linear}. 
The complex field, the sheet-current density in the film, 
and the magnetic flux around the linear wire are given by 
Eqs.~\eqref{cH_linear}, \eqref{Kz_linear}, and \eqref{Phi-wire_linear}, 
respectively, except that now $I_0$ in those equations is replaced 
by $I_0\cos\omega t$. 

The magnetic flux per unit length around the linear wire is given 
by $\Phi_w(t)= L_0I_0\cos\omega t$. 
The voltage induced in the wire defined by Eq.~\eqref{V-wire_def} 
is thus given by $V(t)= R_wI_0\cos\omega t +\omega L_0I_0\sin\omega t$. 
For $0<I_0\le I_{c0}$, the harmonic voltages, the magnetic flux 
per unit length penetrating the film $\Phi_f$, and the electric field 
in the film $E_f$ are all zero.~\cite{foot0}

\subsubsection{Nonlinear response for $I_0>I_{c0}$%===========================
\label{Sec_AC-nonlinear}} For $I_0>I_{c0}$, on the other hand, 
the magnetic field penetrates through the superconducting film, 
as discussed in Sec.~\ref{Sec_DC-nonlinear}. 

For ac drive current $I_0\cos\omega t$, the magnetic flux 
per unit length around the linear wire $\Phi_w(t)$ is~\cite{Zhu93} 
\begin{equation}
	\Phi_w(t)= \widetilde{\Phi}_w(I_0) 
		-2\widetilde{\Phi}_w\Bigl( I_0(1-\cos\omega t)/2 \Bigr) 
\label{Phi_AC-nonlinear}
\end{equation}
for $0<\omega t<\pi$, and $\Phi_w(t)=-\Phi_w(t-\pi/\omega)$ 
for $\pi<\omega t<2\pi$, where $\widetilde{\Phi}_w(I_0)$ is defined 
by Eq.~\eqref{Phi_dc-def}. 
The voltage per unit length of the wire, calculated 
from Eqs.~\eqref{V-wire_def} and \eqref{Phi_AC-nonlinear}, is 
\begin{eqnarray}
	V(t) &=& R_wI_0\cos\omega t 
\nonumber\\
	&& {}+\omega I_0\sin\omega t 
		\widetilde{L}_w\Bigl( I_0(1-\cos\omega t)/2 \Bigr) , 
\label{V_nonlinear}
\end{eqnarray}
where $\widetilde{L}_w(I_0)\equiv d\widetilde{\Phi}_w(I_0)/dI_0$ 
is the differential inductance given by 
\begin{equation}
	\widetilde{L}_w(I_0) = 
	\begin{cases}
		L_0 
		& \mbox{for } 0<I_0<I_{c0} , \\[2ex]
		\displaystyle 
		L_0 +\frac{\mu_0}{2\pi} \ln\left(\frac{I_0}{I_{c0}}\right) 
		& \mbox{for } I_0>I_{c0} . 
	\end{cases}
\label{L_dc-def}
\end{equation}

In response to the ac drive current, the magnetic flux per unit length 
up through the film $\Phi_f(x,t)$ is~\cite{Zhu93} 
\begin{equation}
	\Phi_f(x,t)= \widetilde{\Phi}_f(x,I_0) 
		-2\widetilde{\Phi}_f\Bigl( x,I_0(1-\cos\omega t)/2 \Bigr) 
\label{Phi-film_AC-nonlinear}
\end{equation}
for $0<\omega t<\pi$, and $\Phi_f(x,t)=-\Phi_f(x,t-\pi/\omega)$ 
for $\pi<\omega t<2\pi$, where $\widetilde{\Phi}_f(x,I_0)$ 
is defined by Eq.~\eqref{Phi-film_dc-def}. 
The electric field induced in the film, calculated from 
Eqs.~\eqref{E-film_def} and \eqref{Phi-film_AC-nonlinear}, is 
\begin{equation}
	E_f(x,t)= \omega I_0\sin\omega t 
		\widetilde{L}_f\Bigl( x, I_0(1-\cos\omega t)/2 \Bigr) , 
\label{E-f_nonlinear}
\end{equation}
where the function $\widetilde{L}_f(x,I_0)\equiv 
\partial\widetilde{\Phi}_f(x,I_0)/\partial I_0$ is given by 
\begin{equation}
	\widetilde{L}_f(x,I_0) = 
	\begin{cases}
		0 \qquad\mbox{for } 0<I_0<I_{c0} \mbox{ or } |x|>a , \\[2ex]
		\displaystyle 
		\frac{\mu_0}{2\pi} \mbox{arccosh} \left(\frac{I_0}{I_{c0}} 
		\frac{y_0}{\sqrt{x^2+y_0^2}}\right) \\[2ex]
		\hfill\mbox{for } I_0>I_{c0} \mbox{ and } |x|<a. 
	\end{cases}
\label{L-film_dc-def}
\end{equation}
In order to measure $j_c$ in superconducting films by the inductive 
method detecting the harmonic voltages,~\cite{Claassen91} it is 
important to estimate the induced electric field in superconducting 
films.~\cite{Yamasaki03} The maximum electric field is induced just 
below the linear wire at $x=0$, and is given approximately by 
\begin{equation}
	|E_f|\leq \frac{\mu_0}{\sqrt{2}\pi} \omega I_{c0} 
		\left( \frac{I_0}{I_{c0}}-1 \right) 
% 	|E_f|\leq \mu_0\omega (I_0-I_{c0})/\sqrt{2}\pi 
\label{Ef_max}
\end{equation}
for $0<I_0-I_{c0}\ll I_{c0}$. 

\subsubsection{Harmonic voltages%===========================
\label{Sec_AC-harmonic}}
The voltage per unit length induced in the linear wire can be expressed 
as the Fourier series 
\begin{equation}
	V(t)= \sum_{n=1}^{\infty} V_n\cos(n\omega t+\vartheta_n) , 
\label{V_Fourier}
\end{equation}
where the amplitude $V_n$ and phase difference $\vartheta_n$ 
of the $n$th harmonics are calculated by 
\begin{equation}
	V_n\exp(-i\vartheta_n) = 
	\frac{1}{\pi} \int_0^{2\pi} d(\omega t) V(t)\exp(in\omega t) . 
\label{Vn_def}
\end{equation}
Because of the periodicity of $V(t+\pi/\omega)=-V(t)$, 
the even harmonics are zero; i.e., $V_n=0$ for $n=2,\,4,\,6,\ldots$. 
The fundamental voltage $V_1\exp(-i\vartheta_1)$ is simply determined 
by the resistance per unit length of the linear wire, 
the self-inductance per unit length of the wire, and the mutual 
inductance per unit length between the wire and its image. 
We are interested in the odd harmonics of $V_n\exp(-i\vartheta_n)$ 
for $n=3,\,5,\,7,\ldots$; they are obtained by substituting 
Eqs.~\eqref{V_nonlinear} and \eqref{L_dc-def} into Eq.~\eqref{Vn_def}, 
which yields 
\begin{eqnarray}
	\lefteqn{V_n\exp(-i\vartheta_n)}
\nonumber\\	
	&=& \frac{2}{\pi} \omega I_0 \int_0^{\pi} d\theta \exp(in\theta) 
		\sin\theta \widetilde{L}_w\Bigl( I_0(1-\cos\theta)/2 \Bigr) 
	\hspace{4ex}
\nonumber\\
	&=& \frac{\mu_0}{\pi^2} \omega I_0 
		\int_{\theta_c}^{\pi} d\theta \exp(in\theta) 
		\sin\theta \ln\left( \frac{I_0}{I_{c0}} 
		\frac{1-\cos\theta}{2} \right), 
\label{Vn_odd}
\end{eqnarray}
where $\theta_c=\arccos(1-2I_{c0}/I_0)$. 
For$0<I_0/I_{c0}-1\ll 1$ we have 
\begin{eqnarray}
	V_n &\simeq& \frac{\mu_0}{\pi^2}\omega I_{c0} 
		\left(\frac{I_0}{I_{c0}}-1\right)^2 , 
\label{Vn_I0-Ic0}\\
	\vartheta_n &\simeq& -\pi 
		+\frac{16n}{15} \left(\frac{I_0}{I_{c0}}-1\right)^{1/2} . 
\label{qn_I0-Ic0}
\end{eqnarray}

\begin{figure}[bt]%*************
 \includegraphics{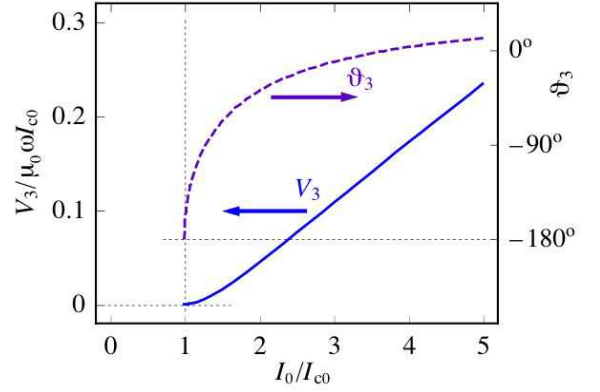}
\caption{(color online) 
The amplitude $V_3$ and the phase $\vartheta_3$ of the third-harmonic 
voltage as functions of the drive current amplitude $I_0$. 
}
\label{Fig_V3-q3-I0}
\end{figure}

Figure~\ref{Fig_V3-q3-I0} shows the $I_0$ dependence of the 
third-harmonic voltage $V_3\exp(-i\vartheta_3)$ calculated from 
Eq.~\eqref{Vn_odd} with $n=3$. 
Although the present model of a linear wire is oversimplified, 
the behavior of the third harmonic voltage shown in this figure 
qualitatively agrees with the experimental data measured by a coil with 
a $\rm YBa_2Cu_3O_{7-y}$ film.~\cite{Claassen91,Poulin93,Mawatari02}

\section{%=====================================================
Two linear wires and a superconducting film
\label{Sec_two-wires}}

\begin{figure}[b]%*************
 \includegraphics{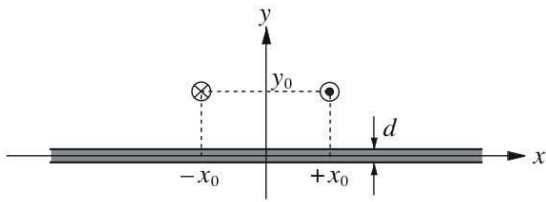}
\caption{%
Configuration of a superconducting film at $y=0$ and two linear wires 
at $(x,y)=(\pm x_0,y_0)$ carrying transport currents $\pm I_z$. 
}
\label{Fig_SC-2wires}
\end{figure}

Two linear wires are better than a single wire to model current-carrying 
coils. 
In this section we consider a superconducting film and two linear wires 
carrying transport currents of opposite directions, as shown in 
Fig.~\ref{Fig_SC-2wires}. 
Infinitely long wires, parallel to the $z$ axis, are situated 
at $(x,y)=(\pm x_0,y_0)$ where $x_0>0$ and $y_0>0$. 
A superconducting film, infinitely extended in the $xz$ plane, 
is situated at $-d/2<y<+d/2$, where $\lambda<d\ll y_0$. 

One wire carrying a dc current $+I_z$ is situated at $(x,y)=(+x_0,y_0)$ 
and another wire carrying a current of opposite direction $-I_z$ is 
at $(x,y)=(-x_0,y_0)$, where $x_0>0$ and $y_0>0$. 

The complex field due to the two wires is given by 
\begin{equation}
	{\cal H}_0(\zeta)= \frac{I_z}{2\pi} \left(\frac{1}{\zeta-\zeta_1} 
		-\frac{1}{\zeta-\zeta_2}\right) , 
\label{h0_2x-wires}
\end{equation}
where $\zeta_1=x_0+iy_0$ and $\zeta_2=-x_0+iy_0=-\zeta_1^*$.

\subsection{DC response%**************************
\label{Sec_DC_2}}
In this subsection we consider the magnetic-field distribution 
when the linear wire carries a dc current $I_z=I_0>0$ 
that is held constant after monotonically increasing from $I_z=0$. 

\subsubsection{Linear response for $0<I_0 \le I_{c0}$%=========================
\label{Sec_DC-linear_2}}
For $0<I_0 \le I_{c0}$, the magnetic field is completely shielded 
below the film, $y=\mbox{Im}\,\zeta <0$. 
The field distribution can be obtained by the mirror-image technique, 
and the resulting complex field is 
\begin{equation}
	{\cal H}(\zeta)= 
	\begin{cases} \displaystyle 
		2{\cal H}_{\parallel}(\zeta) 
		& \mbox{for } \mbox{Im}\,\zeta >0 , \\
		0 & \mbox{for } \mbox{Im}\,\zeta <0 , 
	\end{cases}
\label{cH_linear_2}
\end{equation}
where 
\begin{equation}
	{\cal H}_{\parallel}(\zeta) =\frac{I_0}{2\pi} 
		\left( \frac{\zeta}{\zeta^2-\zeta_1^2} 
		-\frac{\zeta}{\zeta^2-\zeta_2^2} \right) . 
\label{H_para_2}
\end{equation}
The complex potential for $\mbox{Im}\,(\zeta)>0$ is given by 
\begin{equation}
	{\cal G}(\zeta)= \frac{I_0}{2\pi} 
		\ln\left(\frac{\zeta^2-\zeta_1^2}{\zeta^2-\zeta_2^2}\right) . 
\label{G_linear_2}
\end{equation}

The perpendicular magnetic field is $H_y(x,0)=0$, 
and the sheet-current density is thus given by 
\begin{equation}
	K_z(x) = -I_0 F_w(x) 
\label{Kz_linear_2}
\end{equation}
where the function $F_w(x)$ is determined by the configuration of 
the wires, 
\begin{equation}
	F_w(x) = \frac{1}{\pi} \left[ \frac{y_0}{(x-x_0)^2+y_0^2} 
		-\frac{y_0}{(x+x_0)^2+y_0^2}\right] . 
\label{Kz/I0_2wires}
\end{equation}
The sheet-current density $|K_z(x)|$ is maximum at $x=a_c$, where 
\begin{equation}
	a_c= \sqrt{ \frac{1}{3} \left(x_0^2-y_0^2 
		+2\sqrt{x_0^4+x_0^2y_0^2+y_0^4}\right) } . 
\label{ac_2wires}
\end{equation}
The sheet-current density cannot exceed $j_cd$; 
that is, $|K_z(x)|\leq I_0F_w(a_c)\leq j_cd$, 
and Eq.~\eqref{Kz_linear_2} is valid for $I_0\leq I_{c0}$, 
where the threshold current is given by 
\begin{equation}
	I_{c0}= j_c d/F_w(a_c) . 
\label{Ic0_2}
\end{equation}

\begin{figure*}[bt]%*************
 \includegraphics{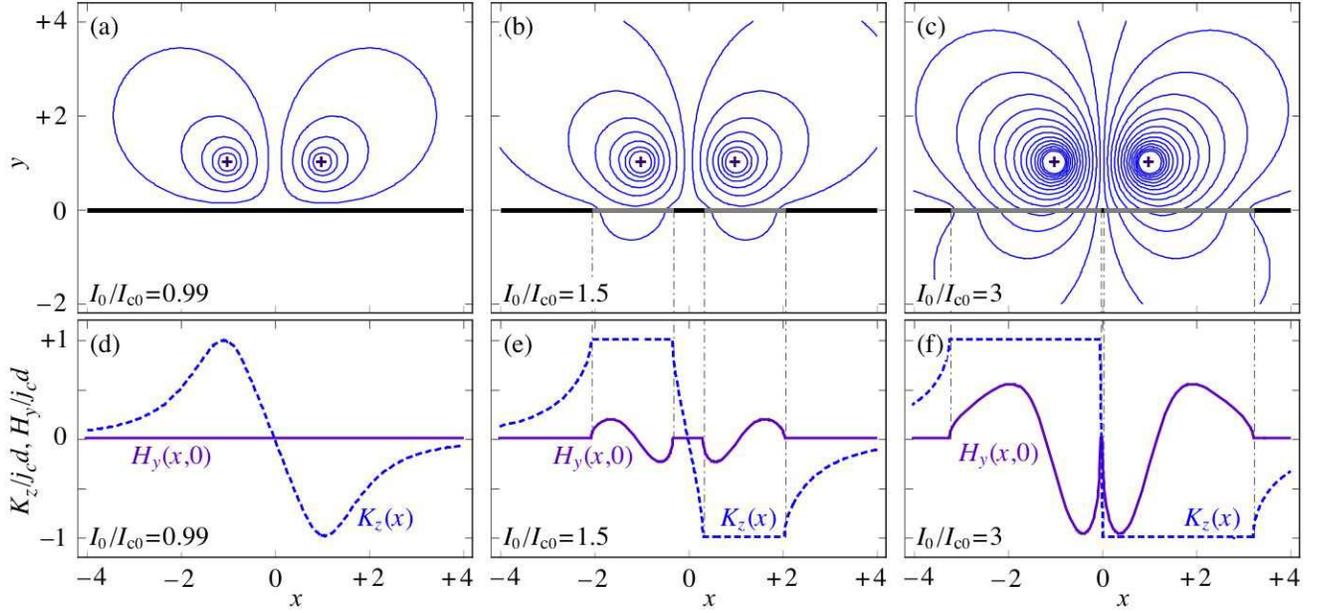}
\caption{(color online) 
Magnetic-field and sheet-current distributions 
(a,d) for $I_0/I_{c0}=0.99$ ($a_c=1.07$), 
(b,e) for $I_0/I_{c0}=1.5$ ($a=2.06$ and $b=0.32$), 
and (c,f) for $I_0/I_{c0}=3$ ($a=3.23$ and $b=0.02$). 
Top figures (a,b,c) show the magnetic-field lines around linear wires 
and a superconducting film. 
The cross symbols at $(x,y)=(\pm x_0,y_0)=(\pm 1,1)$ denote the positions 
of the linear wires, the black horizontal lines the shielded regions 
($|x|<b$ or $|x|>a$) of the superconducting film, and the gray horizontal 
lines the penetrated regions $b<|x|<a$ of the film. 
The vertical dot-dashed lines indicate the positions of the flux fronts 
at $x=\pm a$ and $\pm b$. 
Bottom figures (d,e,f) show the distributions of the perpendicular 
magnetic field $H_y(x,0)$ divided by $j_cd$ (solid lines) 
and the sheet-current density $K_z(x)$ divided by $j_cd$ (dashed lines) 
in the superconducting film. 
}
\label{Fig_field-lines-2wires}
\end{figure*}

Figure~\ref{Fig_field-lines-2wires}(a) shows the magnetic-field lines 
[i.e., the contour lines of $\mbox{Re}\,{\cal G}(x+iy)$] calculated 
from Eq.~\eqref{G_linear_2}, and the dashed line in 
Fig.~\ref{Fig_field-lines-2wires}(d) shows $K_z(x)$ given by 
Eq.~\eqref{Kz_linear_2}. 

%*************

\subsubsection{Nonlinear response for $I_0>I_{c0}$%=========================
\label{Sec_DC-nonlinear_2}}
For $I_0>I_{c0}$, on the other hand, the maximum $|K_z|$ reaches $j_cd$ 
and the magnetic field penetrates below the superconducting film. 
The field distributions for $I_0>I_{c0}$, therefore, must satisfy 
\begin{eqnarray}
	K_z(x) &=& -j_cd\,\mbox{sgn}(x) 
	\quad\mbox{for } b<|x|<a, 
\label{Kz_nonlinear_pntr_2}\\
	H_y(x,0) &=& 0 
	\quad\mbox{for $|x|<b$ or $|x|>a$} , 
\label{Hy_nonlinear_shld_2}
\end{eqnarray}
where the flux fronts are at $x=\pm a$ and $x=\pm b$. 
The complex field and the parameters $a$ and $b$ (where $a>b>0$) are 
determined such that they are consistent with 
Eqs.~\eqref{Kz_nonlinear_pntr_2} and \eqref{Hy_nonlinear_shld_2}, 
as derived in Appendix \ref{App_two-wires-xz}. 

\begin{figure}[b]%*************
 \includegraphics{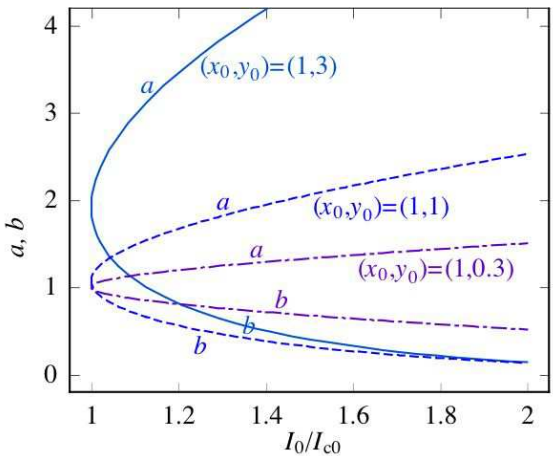}
\caption{(color online) 
The parameters for flux fronts, $a$ (upper lines) and $b$ (lower lines), 
as functions of $I_0$, determined by Eqs.~\eqref{ab-I0_1} and 
\eqref{ab-I0_2}. 
The solid lines show $a$ and $b$ for $(x_0,y_0)=(1,3)$ ($a_c=1.92$), 
the dashed lines for $(x_0,y_0)=(1,1)$ ($a_c=1.07$), 
and the chained lines for $(x_0,y_0)=(1,0.3)$ ($a_c=1.00$). 
}
\label{Fig_2wires-ab-I0}
\end{figure}

The parameters for flux fronts, $a$ and $b$, are determined 
as functions of $I_0$ for $I_0>I_{c0}$, by solving the following 
two equations, 
\begin{eqnarray}
	\frac{I_0}{j_cd}\, \mbox{Re}\, 
		\sqrt{\frac{\zeta_1^2-b^2}{a^2-\zeta_1^2}} 
	&=& a\bm{E}(k) -\frac{b^2}{a}\bm{K}(k) , 
\label{ab-I0_1}\\
	\frac{I_0}{j_cd}\, \mbox{Re}\, 
		\sqrt{\frac{a^2-\zeta_1^2}{\zeta_1^2-b^2}} 
	&=& a[ \bm{K}(k) -\bm{E}(k) ] , 
\label{ab-I0_2}
\end{eqnarray}
where $\bm{K}(k)$ and $\bm{E}(k)$ are the complete elliptic integrals 
of the first and second kinds, and $k=\sqrt{1-b^2/a^2}$. 
These parameters obey $a=b=a_c$ when $I_0=I_{c0}$, and $a>a_c>b>0$ 
when $I_0>I_{c0}$, where $a_c$ is given by Eq.~\eqref{ac_2wires}. 
As shown in Fig.~\ref{Fig_2wires-ab-I0}, $a$ increases 
and $b$ decreases with increasing $I_0$. 

The complex field is derived in Appendix \ref{App_two-wires-xz}, 
and is given by 
\begin{equation}
	{\cal H}(\zeta)= {\cal H}_{\parallel}(\zeta) 
		+{\cal H}_{\perp}(\zeta) +{\cal H}_c(\zeta) , 
\label{H(z)-reduced}
\end{equation}
where 
\begin{eqnarray}
	{\cal H}_{\perp}(\zeta) &=& 
		\frac{I_0}{2\pi} \zeta\phi(\zeta) \left[ 
		\frac{1}{(\zeta^2-\zeta_1^2)\phi(\zeta_1)} 
		-\frac{1}{(\zeta^2-\zeta_2^2)\phi(\zeta_2)} \right] , 
\nonumber\\
\label{Hperp(z)}\\
	{\cal H}_c(\zeta) &=& \frac{j_cd}{\pi} \zeta\phi(\zeta) \int_b^a 
		\frac{du}{(u^2-\zeta^2)\phi(u)} , 
\label{Hc(z)}\\
	\phi(\zeta) &=& \sqrt{(a^2-\zeta^2)(\zeta^2-b^2)} . 
\label{phi(z)}
\end{eqnarray}
Integrating Eq.~\eqref{H(z)-reduced} [or \eqref{H(z)}], we obtain 
the complex potential ${\cal G}(\zeta)=\int{\cal H}(\zeta)d\zeta$, as 
\begin{eqnarray}
	{\cal G}(\zeta) 
	&=& \frac{I_0}{4\pi} 
		\ln\left(\frac{\zeta^2-\zeta_1^2}{\zeta^2-\zeta_2^2}\right) 
\nonumber\\
	&& {}+\frac{I_0}{2\pi} 
		\left[-G_0(\zeta,\zeta_1) +G_0(\zeta,\zeta_2)\right] 
\nonumber\\
	&& {}+\frac{j_cd}{\pi}\int_b^a du G_0(\zeta,u) , 
\label{G(z)}
\end{eqnarray}
where 
\begin{equation}
	G_0(\zeta,s)= \mbox{arctanh} 
		\sqrt{\frac{\zeta^2-b^2}{a^2-\zeta^2} \frac{a^2-s^2}{s^2-b^2}} . 
\label{G0(z)}
\end{equation}

Figure~\ref{Fig_field-lines-2wires}(b) and (c) show the magnetic-field 
lines, i.e., the contour lines of $\mbox{Re}\,{\cal G}(x+iy)$ calculated 
from Eq.~\eqref{G(z)}. 
Figure~\ref{Fig_field-lines-2wires}(e) and (f) show 
$H_y(x,0)=\mbox{Re}\,{\cal H}(x\pm i\epsilon)$ and 
$K_z(x)= \mbox{Im}\,[{\cal H}(x-i\epsilon) -{\cal H}(x+i\epsilon)]$ 
calculated from Eq.~\eqref{H(z)-reduced} 
[or from Eqs.~\eqref{Kz_nonlinear_pntr_2}, \eqref{Hy_nonlinear_shld_2}, 
\eqref{Kz_nonlinear_shld_2}, and \eqref{Hy_nonlinear_pntr_2}]. 

\subsection{AC response%**************************
\label{Sec_AC_2}}

The response of a superconducting film to ac drive currents 
$\pm I_z(t)=\pm I_0\cos\omega t$ flowing in the two linear wires 
at $(x,y)=(\pm x_0,y_0)$ is calculated in a manner similar to 
that described in Sec.~\ref{Sec_AC}. 
The magnetic flux per unit length $\Phi_w$ linked in the two wires 
is calculated from 
\begin{eqnarray}
	\Phi_w &=& -\mu_0\int_{-x_0+r_0}^{+x_0-r_0} dx\,H_y(x,y_0) 
\nonumber\\
	&=& \mu_0 \mbox{Re}\,\Bigl[\, {\cal G}(\zeta_2+r_0) 
		-{\cal G}(\zeta_1-r_0) \,\Bigr] , 
\label{Phi-wire_def-2}
\end{eqnarray}
and the magnetic flux per unit length $\Phi_f(x')$ up through the 
film ($y=0$) in the region $x'<x<+\infty$ is calculated from 
Eq.~\eqref{Phi-film_def}, where ${\cal G}(\zeta)$ is given by 
Eqs.~\eqref{G_linear_2} and \eqref{G(z)}. 
The time-dependent magnetic flux is given by 
Eqs.~\eqref{Phi_AC-nonlinear} and \eqref{Phi-film_AC-nonlinear}. 
The voltage $V(t)$ induced in the wires per unit length and 
the electric field induced in the film are given by 
Eqs.~\eqref{V-wire_def} and \eqref{E-film_def}, respectively.

\section{Discussion%
\label{Sec_discussion}}%************************** 
Here we discuss the more realistic situation of experiments detecting 
the response of a superconducting film to a circular current-carrying 
coil.~\cite{Claassen91} 
We consider a single-turn coil parallel to a superconducting film. 
When the radius of the coil is $x_0$ and the distance between the coil 
and the film is $y_0$, the configuration of the coil and the film 
in the $xy$ plane is similar to that in Fig.~\ref{Fig_SC-2wires}. 
We expect that the magnetic-field lines, the magnetic-field component 
perpendicular to the film, and the circular sheet-current distribution 
in the film then will be similar to the corresponding quantities shown 
in Fig.~\ref{Fig_field-lines-2wires}. 

In actual superconducting films the critical current density $j_c$ 
and the film thickness $d$ can be inhomogeneous, although in the present 
paper we assumed that $j_c d$ is spatially homogeneous. 
The magnetic field produced by the coil of radius $x_0$ is largest directly 
below the coil (i.e., the circular region of radius $\sim x_0$ in the film) 
when $x_0\gtrsim y_0$. 
If $j_c d$ in the circular region below the coil is nearly homogeneous, 
the response of the film will be similar to that presented in 
Sec.~\ref{Sec_two-wires}, and the signal in the coil will yield the local 
value of $j_c d$. 
In this sense the resolution of the measurements using the coil is on the 
order of $x_0$. 
However, if $j_c d$ is very inhomogeneous within the circular region below 
the coil, the response of the film will be quite different from that given 
in Sec.~\ref{Sec_two-wires}.

\section{Conclusion%
\label{Sec_conclusion}}%************************** 
We investigated the response of an infinite superconducting film 
at $y=0$ to linear wires above the film carrying transport currents $I_z$. 
We derived analytic expressions for the complex field, and these are 
given in Eqs.~\eqref{cH_linear}, \eqref{cH_nonlinear}, 
\eqref{cH_linear_2}, and \eqref{H(z)-reduced}. 
The behavior of the film can be summarized as follows. 

(i) Response to a dc current, $I_z=I_0$: 
For a small dc current in the range $0<I_0< I_{c0}\propto j_cd$, 
the sheet-current density $K_z$ in the film is less than the critical 
value $j_cd$, and no magnetic field penetrates into the region $y<0$ 
below the film. 
However, for a large dc current $I_0>I_{c0}$, the sheet-current density 
reaches $|K_z|=j_cd$ and the magnetic field penetrates into 
the region below the film. 

(ii) Response to an ac current, $I_z=I_0\cos\omega t$: 
For $0<I_0<I_{c0}$ the voltage $V(t)$ induced in the linear wire has only 
the first harmonic at the fundamental frequency $\omega_1 = \omega$. 
However, for $I_0>I_{c0}$, odd-harmonic voltages with frequency 
$\omega_n = n\omega$, $n$ = 3, 5, 7,..., are also induced in the linear wire. 

The complex field for the case of a single current-carrying linear wire 
above the film is derived in Appendix A. 
The complex field for the case of a pair of current-carrying wires 
in a plane perpendicular (parallel) to the film is given in Appendix B 
(Appendix C).

\section*{ACKNOWLEDGMENTS} %***** 
We thank H.\ Yamasaki for stimulating discussions. 
This manuscript has been authored in part by Iowa State University of 
Science and Technology under Contract No.\ W-7405-ENG-82 with 
the U.S.\ Department of Energy.

\appendix %*****
\section{\label{App_deriv-cH}% 
Complex field for a superconducting film with 
a linear current-carrying wire above it} %****************** 
\begin{figure}%*************
 \includegraphics{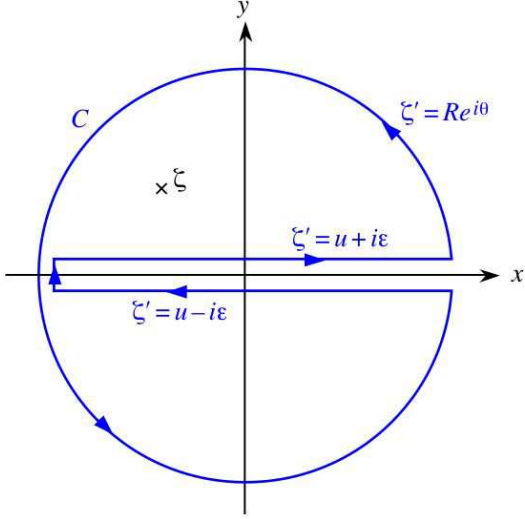}
\caption{(color online) 
The contour $C$ in the $\zeta'$ plane for the integrals 
in Eqs.~\eqref{int-f(z)} and \eqref{int-F(z)} 
consists of a line just above the real axis, 
$\zeta'=u+i\epsilon$ from $u=-R'$ to $u=+R$ where $R'<R$, 
an infinite circle, 
$\zeta'=R e^{i\theta}$ from $\theta=\epsilon/R$ to $\theta=2\pi-\epsilon/R$, 
a line just below the real axis, 
$\zeta'=u-i\epsilon$ from $u=+R$ to $u=-R'$, 
and an infinitesimal line parallel to the imaginary axis, 
$\zeta'=-R'+iv$ from $v=-\epsilon$ to $v=+\epsilon$, 
taking the limit of $R\to\infty$, $R'\to\infty$, and $\epsilon\to+0$. 
}
\label{Fig_contour}
\end{figure}

In this appendix we derive Eqs.~\eqref{a-I0} and \eqref{cH_nonlinear}. 
Consider the function $f(\zeta)$ defined by 
\begin{equation}
	f(\zeta)= \left[{\cal H}(\zeta) -\eta_{\parallel}(\zeta)\right] 
		\sqrt{a^2-\zeta^2} , 
\label{f(z)}
\end{equation}
where $\eta_{\parallel}(\zeta)$ is 
\begin{equation}
	\eta_{\parallel}(\zeta) 
	= \frac{I_0}{2\pi}\frac{iy_0}{\zeta^2+y_0^2} . 
\label{Hpara-1}
\end{equation}
We calculate the integral $\oint_C d\zeta'f(\zeta')/(\zeta'-\zeta)$ 
along the closed contour $C$ shown in Fig.~\ref{Fig_contour}, as 
\begin{equation}
	\frac{1}{2\pi i}\oint_C d\zeta' \frac{f(\zeta')}{\zeta'-\zeta} 
	= \frac{1}{2\pi i}\int_{-\infty}^{+\infty} du 
	\frac{f(u+i\epsilon)-f(u-i\epsilon)}{u-\zeta} , 
\label{int-f(z)}
\end{equation}
where the contour integral along the infinite circle vanishes 
because $|f(\zeta)|\sim |{\cal H}(\zeta)\zeta|\to 0$ for $|\zeta|\to\infty$. 
Because the integrand in the left-hand side of Eq.~\eqref{int-f(z)} 
has poles at $\zeta'=\zeta$ and $\zeta'=\pm iy_0$, the contour integral 
of Eq.~\eqref{int-f(z)} is also calculated by using the residue theorem, 
as $f(\zeta)-f_0(\zeta)$, and thus yielding 
\begin{equation}
	f(\zeta)-f_0(\zeta) =\frac{1}{2\pi i} 
		\int_{-\infty}^{+\infty} du 
		\frac{f(u+i\epsilon)-f(u-i\epsilon)}{u-\zeta} , 
\label{f(z)-int}
\end{equation}
where $f_0(\zeta)$ is given by 
\begin{equation}
	f_0(\zeta)= \frac{I_0}{2\pi} 
		\frac{\zeta\sqrt{a^2+y_0^2}}{\zeta^2+y_0^2} . 
\label{f0(z)}
\end{equation}

For $\zeta=x\pm i\epsilon$, the ${\cal H}(\zeta)$ and 
$\eta_{\parallel}(\zeta)$ are given by 
\begin{eqnarray}
	{\cal H}(x\pm i\epsilon) 
	&=& H_y(x,0) +i\,[\mbox{Im}\,{\cal H}_0(x) \mp K_z(x)/2] , 
	\hspace{3ex}
\label{H(x+i0)}\\
	\eta_{\parallel}(x) &=& i\,\mbox{Im}\,{\cal H}_0(x) , 
\label{Hpara(x)}
\end{eqnarray}
respectively, and thus yielding 
\begin{equation}
	{\cal H}(x\pm i\epsilon)-\eta_{\parallel}(x) 
	= H_y(x,0) \mp iK_z(x)/2 . 
\label{H(x+i0)-Hpara(x)}
\end{equation}
By using Eqs.~\eqref{f(z)}, \eqref{H(x+i0)-Hpara(x)}, and 
\begin{equation}
	\sqrt{a^2-(x\pm i\epsilon)^2}= 
	\begin{cases}
		\mp i\,\mbox{sgn}(x)\sqrt{x^2-a^2} & \mbox{for } |x|>a , \\[1ex]
		\sqrt{a^2-x^2} & \mbox{for } |x|<a , 
	\end{cases}
\end{equation}
we have 
\begin{eqnarray}
	\lefteqn{ f(x+i\epsilon) -f(x-i\epsilon) } 
\nonumber\\
	&=& 
	\begin{cases}
		-2i \mbox{sgn}(x)\sqrt{x^2-a^2} H_y(x,0) 
		& \mbox{for } |x|>a , \\[1ex]
		-i \sqrt{a^2-x^2} K_z(x) 
		& \mbox{for } |x|<a , 
	\end{cases}
\nonumber\\
	&=& 
	\begin{cases}
		0 & \mbox{for } |x|>a , \\[1ex]
		i \sqrt{a^2-x^2} j_cd 
		& \mbox{for } |x|<a , 
	\end{cases}
\nonumber\\
\label{f(x+i0)-f(x-i0)}
\end{eqnarray}
where we used Eqs.~\eqref{KzHy_nonlinear_x<a} and 
\eqref{KzHy_nonlinear_x>a}. 
Substitution of Eq.~\eqref{f(x+i0)-f(x-i0)} into Eq.~\eqref{f(z)-int} 
yields 
\begin{eqnarray}
	f(\zeta)-f_0(\zeta) 
	&=& \frac{j_cd}{2\pi} 
		\int_{-a}^{+a} du \frac{\sqrt{a^2-u^2}}{u-\zeta} , 
\nonumber\\
	&=& \frac{j_cd}{2}\left( -\zeta\pm i\sqrt{a^2-\zeta^2} \right) . 
\label{f(z)-int-calc}
\end{eqnarray}
where the upper (lower) signs hold for $\mbox{Im}\,\zeta >0$ 
($\mbox{Im}\,\zeta <0$). 

From Eqs.~\eqref{f(z)}, \eqref{f0(z)}, and \eqref{f(z)-int-calc} 
we obtain 
\begin{eqnarray}
	{\cal H}(\zeta) &=& 
		\frac{I_0}{2\pi} \frac{1}{\zeta^2+y_0^2} 
		\left(iy_0 +\frac{\zeta\sqrt{a^2+y_0^2}}{%
		\sqrt{a^2-\zeta^2}} \right) 
\nonumber\\
	&& {}+\frac{j_cd}{2} 
		\left( \pm i-\frac{\zeta}{\sqrt{a^2-\zeta^2}} \right) . 
\label{cH_calc-1}
\end{eqnarray}
Note that the first term of the right-hand side of Eq.~\eqref{cH_calc-1}, 
which is proportional to $I_0$, corresponds to the complex field 
in the ideal Meissner state for two semi-infinite strips situated 
at $|\mbox{Re}\,\zeta |>a$ with a line current at $\zeta =iy_0$. 
By using $\sqrt{-\zeta^2}=\mp i\zeta$, we confirm that 
Eq.~\eqref{cH_calc-1} for $a\to 0$ is identical to Eq.~\eqref{cH_linear}. 
Equation~\eqref{cH_calc-1} can be rewritten as 
\begin{eqnarray}
	{\cal H}(\zeta) &=& \pm i\,\frac{j_cd}{2} 
		+\frac{I_0}{2\pi} \frac{1}{\zeta^2+y_0^2} 
		\left(iy_0 +\frac{\zeta\sqrt{a^2-\zeta^2}}{%
		\sqrt{a^2+y_0^2}} \right) 
\nonumber\\
	&& {}+\frac{\zeta}{\sqrt{a^2-\zeta^2}} 
		\left( \frac{I_0}{2\pi}\frac{1}{\sqrt{a^2+y_0^2}} 
		-\frac{j_cd}{2} \right) . 
\label{cH_calc-2}
\end{eqnarray}
However, the last term of the right-hand side of Eq.~\eqref{cH_calc-2} 
must vanish, because ${\cal H}(\zeta)$ is finite at $\zeta=\pm a$. 
We thus obtain $I_0=\pi j_cd\sqrt{a^2+y_0^2}$, 
such that the parameter $a$ describing the position of the flux front 
is given by Eq.~\eqref{a-I0}. 
The resulting expression for the complex field is given by 
Eq.~\eqref{cH_nonlinear}. 

See also Appendix \ref{App_two-wires} for the complex field 
for a superconducting film with two wires in the $yz$ plane.

\section{\label{App_two-wires}
Complex field for two parallel wires in the $yz$ plane}
%****************** 
We present here the complex field for a superconducting film 
with two parallel linear wires above it. 
An infinite superconducting film is in the plane $y=0$, 
as shown in Fig.~\ref{Fig_SC-wire}. 
One wire carrying a dc current $+I_0$ is situated at $(x,y)=(0,y_1)$ 
and another wire carrying current $-I_0$ is at $(x,y)=(0,y_2)$, 
where $0<y_1<y_2$. 
The derivation of the complex field for the two wires is similar 
to that given in Appendix~\ref{App_deriv-cH} for a single wire, 
and we exhibit here only the resulting expressions. 

For $0<I_0<I_{c0}$, the sheet-current density and the perpendicular 
magnetic field satisfy $|K_z(x)|<j_cd$ and $H_y(x,0)=0$, respectively. 
The threshold current is given by 
\begin{equation}
	I_{c0}= \pi j_cd \frac{y_1y_2}{y_2-y_1} , 
\label{Ic0_2wires}
\end{equation}
and the complex field for $0<I_0<I_{c0}$ is 
\begin{equation}
	{\cal H}(\zeta)= 
	\begin{cases} \displaystyle 
		\frac{I_0}{\pi}\left(\frac{iy_1}{\zeta^2+y_1^2} 
		-\frac{iy_2}{\zeta^2+y_2^2}\right) 
		& \mbox{for } \mbox{Im}\,\zeta >0 , \\
		0 & \mbox{for } \mbox{Im}\,\zeta <0 . 
	\end{cases}
\label{cH_linear-2wires}
\end{equation}
For $I_0>I_{c0}$, $K_z(x)$ and $H_y(x,0)$ fulfill 
Eqs.~\eqref{KzHy_nonlinear_x<a} and \eqref{KzHy_nonlinear_x>a}, 
and the relationship between $a$ and $I_0$ is given by 
\begin{equation}
	I_0= \pi j_cd \left(\frac{1}{\sqrt{a^2+y_1^2}} 
	-\frac{1}{\sqrt{a^2+y_2^2}}\right)^{-1} . 
\label{a-I0_two-wires}
\end{equation}
% \begin{widetext} 
The complex field and complex potential for $I_0>I_{c0}$ are 
respectively given by 
\begin{eqnarray}
	{\cal H}(\zeta) &=& \pm i\, \frac{j_cd}{2} 
		+\frac{I_0}{2\pi} \left[ \frac{1}{\zeta^2+y_1^2} 
		\left(iy_1 +\frac{\zeta\sqrt{a^2-\zeta^2}}{%
		\sqrt{a^2+y_1^2}} \right)\right. 
\nonumber\\
	&& \left. {}-\frac{1}{\zeta^2+y_2^2} 
		\left(iy_2 +\frac{\zeta\sqrt{a^2-\zeta^2}}{%
		\sqrt{a^2+y_2^2}} \right)\right] , 
\label{cH_two-wires}\\
	{\cal G}(\zeta) &=& \frac{j_cd}{2} 
		\left(\pm i\zeta +\sqrt{a^2-\zeta^2} \right) 
\nonumber\\
		&& {}+\frac{I_0}{2\pi} 
		\left[ i\arctan\left(\frac{\zeta}{y_1}\right) 
		-\mbox{arctanh}\left(\sqrt{%
		\frac{a^2-\zeta^2}{a^2+y_1^2}}\right) 
		\right. \nonumber\\ && \left. {}
		-i\arctan\left(\frac{\zeta}{y_2}\right) 
		+\mbox{arctanh}\left(\sqrt{%
		\frac{a^2-\zeta^2}{a^2+y_2^2}}\right) \right] . 
\nonumber\\
\label{cG_two-wires}
\end{eqnarray}

\section{\label{App_two-wires-xz}
Complex field for two parallel wires in a plane parallel to the $xz$ plane}
%****************** 
In this appendix we derive Eqs.~\eqref{ab-I0_1}--\eqref{Hc(z)}. 
Consider the function $F(\zeta)$ defined by 
\begin{equation}
	F(\zeta)= \left[{\cal H}(\zeta)-{\cal H}_{\parallel}(\zeta)\right] 
		\phi(\zeta) , 
\label{F(z)}
\end{equation}
where ${\cal H}_{\parallel}(\zeta)$ and $\phi(\zeta)$ are defined by 
Eqs.~\eqref{H_para_2} and \eqref{phi(z)}, respectively. 
We calculate the integral $\oint_C d\zeta'F(\zeta')/(\zeta'-\zeta)$ 
along the closed contour $C$ shown in Fig.~\ref{Fig_contour}, 
as in Eq.~\eqref{int-f(z)}, 
\begin{equation}
	\frac{1}{2\pi i}\oint_C d\zeta' \frac{F(\zeta')}{\zeta'-\zeta} 
	= \frac{1}{2\pi i}\int_{-\infty}^{+\infty} du 
	\frac{F(u+i\epsilon)-F(u-i\epsilon)}{u-\zeta} , 
\label{int-F(z)}
\end{equation}
where the contour integral along the infinite circle vanishes because 
$|F(\zeta)|\sim |{\cal H}(\zeta)\zeta^2|\to 0$ for $|\zeta|\to\infty$. 
Because the integrand on the left-hand side of Eq.~\eqref{int-F(z)} 
has poles at $\zeta'=\zeta$, $\pm\zeta_1$, and $\pm\zeta_2$, the contour 
integral of Eq.~\eqref{int-F(z)} can be calculated by using the residue 
theorem, as $F(\zeta)-F_0(\zeta)$, and thus yielding 
\begin{equation}
	F(\zeta)-F_0(\zeta) =\frac{1}{2\pi i} 
		\int_{-\infty}^{+\infty} du 
		\frac{F(u+i\epsilon)-F(u-i\epsilon)}{u-\zeta} . 
\label{F(z)-int}
\end{equation}
The $F_0(\zeta)$ is defined by 
\begin{eqnarray}
	F_0(\zeta) &=& \frac{I_0}{4\pi} 
		\left[ \frac{\phi(\zeta_1)}{\zeta-\zeta_1} 
		-\frac{\phi(\zeta_2)}{\zeta-\zeta_2} 
		-\frac{\phi(-\zeta_1)}{\zeta+\zeta_1} 
		+\frac{\phi(-\zeta_2)}{\zeta+\zeta_2} \right] 
\nonumber\\
	&=& \frac{I_0}{2\pi} 
		\left[ \frac{\zeta\phi(\zeta_1)}{\zeta^2-\zeta_1^2} 
		-\frac{\zeta\phi(\zeta_2)}{\zeta^2-\zeta_2^2} \right] , 
\label{F0(z)}
\end{eqnarray}
where we used $\phi(-\zeta_1)=-\phi(\zeta_1)$ and 
$\phi(-\zeta_2)=-\phi(\zeta_2)$. 

For $\zeta=x\pm i\epsilon$, Eq.~\eqref{phi(z)} is reduced to 
\begin{equation}
	\phi(x\pm i\epsilon)= 
	\begin{cases}
		\pm i\widetilde{\phi}(x) & \mbox{for $|x|< b$ or $|x|>a$} , \\[1ex]
		\mbox{sgn}(x) \widetilde{\phi}(x) & \mbox{for $b<|x|<a$} , 
	\end{cases}
\label{phi(x+i0)}
\end{equation}
where 
\begin{eqnarray}
	\widetilde{\phi}(x) 
	&=& \mbox{sgn}(a-|x|) \sqrt{\left|(a^2-x^2)(x^2-b^2)\right|} 
\label{tilde-phi(x)}\\
	&=& \begin{cases}
		\sqrt{(a^2-x^2)(b^2-x^2)} & \mbox{for $|x|< b$} , \\[1ex]
		\sqrt{(a^2-x^2)(x^2-b^2)} & \mbox{for $b<|x|<a$} , \\[1ex]
		-\sqrt{(x^2-a^2)(x^2-b^2)} & \mbox{for $|x|>a$} . 
	\end{cases}
\nonumber
\end{eqnarray}
Substituting Eqs.~\eqref{H(x+i0)-Hpara(x)} and \eqref{phi(x+i0)} 
into Eq.~\eqref{F(z)} with $\zeta=x\pm i\epsilon$, we have 
\begin{eqnarray}
	\lefteqn{F(x+i\epsilon)-F(x-i\epsilon)} 
\nonumber\\
	&=& \begin{cases}
		2i\widetilde{\phi}(x) H_y(x,0) 
		& \mbox{for $|x|< b$ or $|x|>a$} , \\[1ex]
		-i\,\mbox{sgn}(x) \widetilde{\phi}(x) K_z(x) 
		& \mbox{for $b<|x|<a$} , 
		\end{cases}
\nonumber\\
	&=& \begin{cases}
		0 
		& \mbox{for $|x|< b$ or $|x|>a$} , \\[1ex]
		i\,j_cd\, \widetilde{\phi}(x) 
		& \mbox{for $b<|x|<a$} , 
		\end{cases}
\label{F(x+i0)-F(x-i0)}
\nonumber\\
\end{eqnarray}
where we used Eqs.~\eqref{Kz_nonlinear_pntr_2} and 
\eqref{Hy_nonlinear_shld_2}. 
Substitution of Eq.~\eqref{F(x+i0)-F(x-i0)} into Eq.~\eqref{F(z)-int} 
yields 
\begin{eqnarray}
	F(\zeta) -F_0(\zeta) &=& \frac{j_cd}{2\pi} 
		\int_{b<|u|<a} du \frac{\widetilde{\phi}(u)}{u-\zeta} , 
\nonumber\\
	&=& \frac{j_cd}{\pi} 
		\int_b^a du \frac{\zeta\phi(u)}{u^2-\zeta^2} , 
\label{F(z)-int-2}
\end{eqnarray}
such that 
\begin{equation}
	{\cal H}(\zeta)= {\cal H}_{\parallel}(\zeta) 
		+\frac{F_0(\zeta)}{\phi(\zeta)} 
		+\frac{j_cd}{\pi} \frac{\zeta}{\phi(\zeta)}
		\int_b^a du \frac{\phi(u)}{u^2-\zeta^2} . 
\label{H(z)}
\end{equation}
By using 
\begin{equation}
	\frac{\phi(s)}{\phi(\zeta)} 
	= \frac{\phi(\zeta)}{\phi(s)} 
		+\frac{\left(\zeta^2-s^2\right) 
		\left(\zeta^2+s^2-a^2-b^2\right)}{\phi(\zeta) \phi(s)} , 
\label{phi-phi}
\end{equation}
we rewrite Eq.~\eqref{H(z)} as 
\begin{eqnarray}
	\lefteqn{ {\cal H}(\zeta) -{\cal H}_{\parallel}(\zeta) }
\nonumber\\
	&=& \zeta\phi(\zeta)\left\{ \frac{I_0}{2\pi}\left[ 
		\frac{1}{(\zeta^2-\zeta_1^2)\phi(\zeta_1)} 
		-\frac{1}{(\zeta^2-\zeta_2^2)\phi(\zeta_2)} \right] \right.
\nonumber\\
	&& \left.\quad{}+\frac{j_cd}{\pi} \int_b^a 
		\frac{du}{(u^2-\zeta^2)\phi(u)} \right\} 
\nonumber\\
	&& {}+\frac{\zeta}{\phi(\zeta)}{\cal R}(\zeta) , 
\label{H(z)-calc}
\end{eqnarray}
where 
\begin{eqnarray}
	{\cal R}(\zeta) &=& \frac{I_0}{2\pi} 
		\left[ \frac{\zeta^2+\zeta_1^2-a^2-b^2}{\phi(\zeta_1)} 
		-\frac{\zeta^2+\zeta_2^2-a^2-b^2}{\phi(\zeta_2)} \right] 
\nonumber\\
	&& {}-\frac{j_cd}{\pi}\int_b^a du 
		\frac{\zeta^2+u^2-a^2-b^2}{\widetilde\phi(u)} 
\nonumber\\
	&=& \frac{(\zeta^2-b^2){\cal R}(\pm a) 
		+(a^2-\zeta^2){\cal R}(\pm b)}{a^2-b^2} . 
\label{R(z)}
\end{eqnarray}
For $\zeta=\pm a$ and $\pm b$, we have 
\begin{eqnarray}
	{\cal R}(\pm a) &=& \frac{I_0}{2\pi} 
		\left( \sqrt{\frac{\zeta_1^2-b^2}{a^2-\zeta_1^2}} 
		-\sqrt{\frac{\zeta_2^2-b^2}{a^2-\zeta_2^2}} \right) 
\nonumber\\
	&& {}-\frac{j_cd}{\pi}\int_b^a du 
		\sqrt{\frac{u^2-b^2}{a^2-u^2}} , 
\label{R(a)}\\
	{\cal R}(\pm b) &=& \frac{I_0}{2\pi} 
		\left( \sqrt{\frac{a^2-\zeta_1^2}{\zeta_1^2-b^2}} 
		-\sqrt{\frac{a^2-\zeta_2^2}{\zeta_2^2-b^2}} \right) 
\nonumber\\
	&& {}-\frac{j_cd}{\pi}\int_b^a du 
		\sqrt{\frac{a^2-u^2}{u^2-b^2}} . 
\label{R(b)}
\end{eqnarray}
In order to remove the divergences in Eq.~\eqref{H(z)-calc} 
at $\zeta=\pm a$ and $\pm b$, we require 
${\cal R}(\pm a)={\cal R}(\pm b)=0$, thus yielding 
\begin{eqnarray}
	\frac{I_0}{j_cd} \mbox{Re}\,\sqrt{\frac{\zeta_1^2-b^2}{a^2-\zeta_1^2}} 
	&=& \int_b^a du\sqrt{\frac{u^2-b^2}{a^2-u^2}} , 
\label{ab-1}\\
	\frac{I_0}{j_cd} \mbox{Re}\,\sqrt{\frac{a^2-\zeta_1^2}{\zeta_1^2-b^2}} 
	&=& \int_b^a du\sqrt{\frac{a^2-u^2}{u^2-b^2}} . 
\label{ab-2}
\end{eqnarray}
Equations~\eqref{ab-1} and \eqref{ab-2} reduce to Eqs.~\eqref{ab-I0_1} 
and \eqref{ab-I0_2}, respectively. 
We thus obtain ${\cal R}(\zeta)=0$ for any $\zeta$, 
and Eq.~\eqref{H(z)-calc} is reduced to Eqs.~\eqref{H(z)-reduced}, 
\eqref{Hperp(z)}, and \eqref{Hc(z)}.

%\subsection{Sheet Current and Perpendicular Magnetic Field} %*****
The sheet current $K_z(x)=\mbox{Im}\,[{\cal H}(x-i\epsilon) 
-{\cal H}(x+i\epsilon)]$ and the perpendicular magnetic field 
$H_y(x,0)=\mbox{Re}\,{\cal H}(x\pm i\epsilon)$ are obtained by 
substituting $\zeta=x\pm i\epsilon$ in Eq.~\eqref{H(z)-reduced} 
with Eq.~\eqref{H(x+i0)-Hpara(x)}, 
\begin{eqnarray}
	\lefteqn{ H_y(x,0)\mp i\,K_z(x)/2 
	= {\cal H}(x\pm i\epsilon)-{\cal H}_{\parallel}(x) }
\nonumber\\
	&=& \phi(x\pm i\epsilon) \left\{ \frac{I_0}{2\pi} 
		\left[ \frac{x}{(x^2-\zeta_1^2)\phi(\zeta_1)} 
		-\frac{x}{(x^2-\zeta_2^2)\phi(\zeta_2)} \right]\right. 
\nonumber\\
	&& \left.{}-\frac{j_cd}{\pi}\int_b^a 
		\frac{du}{[(x\pm i\epsilon)^2-u^2]\phi(u)} \right\} 
\nonumber\\
	&=& \phi(x\pm i\epsilon) \left\{ \frac{I_0}{\pi} 
		\mbox{Re}\,\left[ \frac{x}{(x^2-\zeta_1^2)\phi(\zeta_1)} 
		\right]\right. 
\nonumber\\
	&& {}-\frac{j_cd}{\pi}\int_b^a \frac{du}{\phi(u)} 
		\left[\frac{x}{x^2-u^2} \right.
\nonumber\\
	&& \left.\left.\quad{}\mp\frac{i\pi}{2} 
		\Bigl(\delta(x-u)+\delta(x+u)\Bigr)\right] \right\} . 
\label{HyKz-cal}
\end{eqnarray}
Substitution of Eq.~\eqref{phi(x+i0)} into Eq.~\eqref{HyKz-cal} yields 
the sheet current $K_z(x)$ and the perpendicular magnetic field $H_y(x,0)$. 
The $K_z(x)$ for $b<|x|<a$ is given by Eq.~\eqref{Kz_nonlinear_pntr_2}, 
and that for $|x|<b$ or $|x|>a$ is given by 
\begin{eqnarray}
	K_z(x) &=& \frac{2}{\pi} x\widetilde{\phi}(x) 
		\left\{ I_0 \mbox{Re}\,\left[ 
		\frac{1}{(x^2-\zeta_1^2)\phi(\zeta_1)} \right] \right.
\nonumber\\
	&& \left. {}-j_cd \int_{b}^{a} 
		\frac{du}{(x^2-u^2)\phi(u)} \right\} . 
\label{Kz_nonlinear_shld_2}
\end{eqnarray}
The $H_y(x,0)$ for $|x|<b$ or $|x|>a$ is given by 
Eq.~\eqref{Hy_nonlinear_shld_2}, and that for $b<|x|<a$ is given by 
\begin{eqnarray}
	H_y(x,0) &=& \frac{1}{\pi}|x| \widetilde{\phi}(x) 
		\left\{ I_0 \mbox{Re}\,\left[ 
		\frac{1}{(x^2-\zeta_1^2)\phi(\zeta_1)} \right] \right. 
\nonumber\\
	&& \left. {}-j_cd\, \mbox{P} \int_{b}^{a} 
		\frac{du}{(x^2-u^2)\phi(u)} \right\} , 
\label{Hy_nonlinear_pntr_2}
\end{eqnarray}
where P denotes the Cauchy principal value integral.


\begin{thebibliography}{99}%*****
\bibitem{Swan68}
G. W. Swan, J. Math. Phys. {\bf 9}, 1308 (1968).
\bibitem{Halse70}
M. R. Halse, J. Phys. D {\bf 3}, 717 (1970).
\bibitem{Mikheenko93}
P. N. Mikheenko and Yu. E. Kuzovlev, Physica C {\bf 204}, 229 (1993).
\bibitem{Zhu93}
J. Zhu, J. Mester, J. Lockhart and J. Turneaure, 
Physica C {\bf 212}, 216 (1993).
\bibitem{Brandt93}
E. H. Brandt, M. V. Indenbom and A. Forkl, 
Europhys Lett. {\bf 22}, 735 (1993); 
% \bibitem{Brandt93b} 
E. H. Brandt and M. Indenbom, 
\prb {\bf 48}, 12893 (1993).
\bibitem{Zeldov94a}
E. Zeldov, J. R. Clem, M. McElfresh and M. Darwin, 
\prb {\bf 49}, 9802 (1994).
\bibitem{Bean62}
C. P. Bean, \prl {\bf 8}, 250 (1962); \rmp {\bf 36}, 31 (1964).

\bibitem{Fiory88}
A. T. Fiory, A. F. Hebard, P. M. Mankiewich, and R. E. Howard, 
\apl {\bf 52}, 2165 (1988).
\bibitem{Clem92}
J. R. Clem and M. W. Coffey, \prb {\bf 46}, 14662 (1992).
\bibitem{Klupsch95}
Th. Klupsch and M. Zeisberger, Physica C {\bf 244}, 153 (1995).
\bibitem{Gilchrist96}
J. Gilchrist and E. H. Brandt, \prb {\bf 54}, 3530 (1996).
\bibitem{Turneaure98}
S. J. Turneaure, A. A. Pesetski, and T. R. Lemberger, 
J. Appl. Phys. {\bf 83}, 4334, (1998).
\bibitem{Coffey01}
M. W. Coffey, J. Appl. Phys. {\bf 89}, 5570 (2001).

\bibitem{Bernhardt89}
K. Bernhardt, R. Gross, M. Hartmann, R. P. Huebener, F. Kober, 
D. Koelle, and T. Sermet, 
Physica C {\bf 161}, 468 (1989).
\bibitem{Claassen91}
J. H. Claassen, M. E. Reeves, and R. J. Soulen, Jr., 
Rev. Sci. Instrum. {\bf 62}, 996 (1991).
\bibitem{Poulin93}
G. D. Poulin, J. S. Preston, and T. Strach, \prb {\bf 48}, 1077 (1993).
\bibitem{Hochmuth94}
H. Hochmuth and M. Lorenz, Physica C {\bf 220}, 209 (1994); 
ibid. {\bf 265}, 335 (1996).
\bibitem{Mawatari02}
Y. Mawatari, H. Yamasaki, and Y. Nakagawa, \apl {\bf 81}, 2424 (2002).

\bibitem{Koo96}
L. S. Koo and K. L. Telschow, \prb {\bf 53}, 8743 (1996).
\bibitem{Wada03}
H. Wada, M. Migita, E. S. Otabe, M. Kiuchi, T. Matsushita, 
Y. Mawatari, H. Yamasaki, and Y. Nakagawa, 
Physica C {\bf 392-396}, 1310 (2003); 
T. Nadami, E. S. Otabe, M. Kiuchi, and T. Matsushita, 
ibid. {\bf 412-414}, 1011 (2004); 
T. Nadami, E. S. Otabe, M. Kiuchi, T. Matsushita, 
Y. Mawatari, H. Yamasaki, and Y. Nakagawa, 
ibid. {\bf 426-431}, 688 (2005).
\bibitem{Yamada05}
H. Yamada, A. Bitoh, Y. Mitsuno, I. Imai, K. Nomura, K. Kanayama, 
S. Nakagawa, Y. Mawatari, and H. Yamasaki, 
Physica C {\bf 433}, 59 (2005).
\bibitem{Aurino05}
M. Aurino, E. Di Gennaro, F. Di Iorio, A. Gauzzi, G. Lamura, 
and A. Andreone, 
J. Appl. Phys. {\bf 98}, 123901 (2005).

\bibitem{Yamasaki03}
H. Yamasaki, Y. Mawatari, and Y. Nakagawa, \apl {\bf 82}, 3275 (2003).

\bibitem{Clem06}
The flux-penetration process for finite $H_{c1}/j_cd>0$ is more 
complicated than that shown in the present paper for $H_{c1}/j_cd\to 0$. 
The case for $H_{c1}/j_cd>0$ will be published elsewhere, 
J. R. Clem and Y. Mawatari, unpublished. 

\bibitem{Clem73}
J. R. Clem, R. P. Huebener, and D. E. Gallus, 
J. Low Temp. Phys. {\bf 12}, 449 (1973).
\bibitem{Zeldov94b}
E. Zeldov, A. I. Larkin, V. B. Geshkenbein, M. Konczykowski, 
D. Majer, B. Khaykovich, V. M. Vinokur, and H. Shtrikman, 
\prl {\bf 73}, 1428 (1994).
\bibitem{Mawatari01}
Y. Mawatari and J. R. Clem, Phys. Rev. Lett. {\bf 86}, 2870 (2001); 
%\bibitem{Brojeny02}
A. A. Babaei Brojeny, Y. Mawatari, M. Benkraouda, and J. R. Clem, 
Supercond. Sci. Technol. {\bf 15}, 1454 (2002); 
%\bibitem{Mawatari03}
Y. Mawatari and J. R. Clem, \prb {\bf 68}, 024505 (2003).

\bibitem{foot0}
Note that the simple result of $\Phi_f=E_f=0$ for $0<I_0<I_{c0}$ 
is ideally accurate only for the thin-film limit, $d\to 0$. 
Small but finite magnitudes of $|\Phi_f|$ and $|E_f|$ arise 
even for $0<I_0<I_{c0}$ in a superconducting film of finite thickness, 
because the magnetic flux penetrates in the vicinity of the upper 
surface of the film. 

\bibitem{foot1}
We confine our attention here to low frequencies, well below those 
normally used to investigate the surface resistance of superconducting 
films, where energy dissipation arises due to excitation of 
quasiparticles or to small-amplitude vortex motion, as discussed, 
for example, in M. W. Coffey and J. R. Clem, \prl {\bf 67}, 386 (1991). 

\end{thebibliography}
\end{document}